\documentclass{Rinton-P10x7}

\begin{document}

\title{Conditional quantum evolution induced by continuous measurement
for a mesoscopic qubit}

\author{Hsi-Sheng Goan and  Gerard J.~Milburn}
\address{Center for Quantum Computer Technology and Department of
Physics, University of Queensland, Brisbane, Qld 4072, Australia}
%${}^2$School of Science, Griffith University, Nathan,
%Qld 4111 Australia}


\maketitle

\abstracts{We consider the problem of an electron tunneling between 
two coupled quantum dots, a two-state quantum system (qubit),
using a low-transparency point contact (PC) or tunnel junction
as a detector continually measuring the position of the electron.
We focus on the qubit dynamics conditioned on a particular realization
of the actual measured current through the PC device.
We illustrate the conditional evolutions by numerical simulations.
The different behaviors between unconditional and
conditional evolutions are demonstrated.
The conditional qubit dynamics evolving
from quantum jumps to quantum diffusion is presented.}

%%%%%%%%%%%%%%%%%%%%%%%%%%%%%%%%%%%%%%%%%%%%%%%%%%%%%%%%%%
\section{Introduction}
%%%%%%%%%%%%%%%%%%%%%%%%%%%%%%%%%%%%%%%%%%%%%%%%%%%%%%%%%%

In condensed matter physics, usually many identical quantum systems are 
prepared at the same time when a measurement is made upon the systems.
For example, in nuclear or electron magnetic resonance experiments, 
generally an ensemble of systems of nuclei and 
electrons are probed to obtain the resonance signals. 
This implies that the measurement result in this case is 
an average response of the ensemble. 
On the other hand, for various proposed condensed-matter quantum computer
architectures, 
the need to readout physical properties of a single electronic qubit, 
such as charge or spin at a single electron level, is demanding. 
It is particularly important to take account 
of the decoherence introduced by the measurements on the qubit 
as well as to understand how the quantum state of the qubit, 
conditioned on a particular single realization of measurement, 
evolves in time for the purpose of quantum computing.

We consider, in this paper, the
problem of an electron tunneling between two coupled quantum dots
(CQDs), a two-state quantum system (qubit),
using a low-transparency point contact (PC) or tunnel junction
as a detector (environment) continuously measuring the position
of the electron. 
%schematically illustrated in Fig.\ \ref{fig:PC}. 
This problem has been extensively studied in Refs.\
\cite{Gurvitz97,Korotkov99,Makhlin98,Goan00}.
The master equation (or rate equations
for all the reduced-density-matrix elements) 
for the CQD system (qubit) 
has been
derived and analyzed 
in Refs.\ \cite{Gurvitz97,Goan00}. 
This (unconditional) master equation
is obtained when the results of all
measurement records
(electron current records in this case)
are completely ignored or averaged over, and
describes only the ensemble average property
for the qubit.
However, if a measurement
is made on the system and the results are available, the state or density
matrix is a
conditional state conditioned on the measurement results.
Hence the deterministic, unconditional master equation cannot
describe the conditional dynamics of the qubit in a single realization
of continuous measurements which reflects the stochastic nature of an
electron tunneling through the PC barrier.

Korotkov \cite{Korotkov99} has obtained the Langevin rate
equations for the CQD system measured by a ideal PC detector
in the quantum-diffusive limit\cite{Goan00}.
These rate equations describe the random
evolution of the density matrix that both conditions, and is conditioned
by, the
PC detector output.
Recently, we \cite{Goan00}
presented a {\em quantum trajectory}
measurement analysis of the same system.
We found that the conditional dynamics of the 
qubit can be described by the
stochastic Schr\"{o}dinger equation
for the conditioned state vector, provided that the information
carried away from the qubit by the PC reservoirs can be recovered by
the perfect detection of the measurements.
We also analyzed the localization rates at which the
qubit becomes localized in one of the two states 
when the coupling frequency $\Omega$ between the states is zero.
We showed that the localization time discussed there
is slightly different from
the measurement time defined in
Refs.\ \cite{Makhlin98}.
The mixing rate at which the two possible states of
the qubit become mixed when $\Omega\neq 0$ was calculated as well
and found in agreement with
the result in Ref.\ \cite{Makhlin98}.
In this paper,
we focus on the qubit dynamics conditioned on a particular realization
of the actual measured current through the PC device.
We illustrate the conditional evolutions by numerical simulations.
The different behavior between unconditional and
conditional evolutions are demonstrated.
Furthermore, the conditional qubit dynamics evolving
from quantum jumps to quantum diffusion \cite{Goan00} is presented.

%%%%%%%%%%%%%%%%%%%%%%%%%%%%%%%%%%%%%%%%%%%%%%%%%%%%%%%%%%
\section{Conditional quantum evolution under continuous measurement}
\label{sec:analytic}
%%%%%%%%%%%%%%%%%%%%%%%%%%%%%%%%%%%%%%%%%%%%%%%%%%%%%%%%%%

The whole CQD and PC model is described in
Refs.\ \cite{Gurvitz97,Korotkov99,Goan00}.
Basically, when the electron
in the CQD system is near the PC (i.e., dot $1$ is occupied),
there is a change in the PC tunneling barrier, 
which then
results in a
change in the effective tunneling amplitude.
As a consequence, the current through the PC is
also modified. This changed current can be detected,
and thus a measurement of 
the location of the electron in the CQD
system (qubit) is effected.
We describe parameters used in this paper
in the following.
We denote the logical qubit states 
as $|a\rangle$ (i.e., dot 1 is occupied)
and $|b\rangle$ (i.e., dot 2 is occupied). 
The coupling and energy mismatch
between the qubit states are $\hbar\Omega$,
and $\hbar {\cal E}$ respectively.
$\Gamma_d={|{\cal X}|^2}/{2}$ is the 
decoherence (dephasing) rate generated by the PC reservoirs
in the unconditional dynamics. 
The parameters ${\cal X}$ and ${\cal T}$
are given by: $|{\cal T}+{\cal X}|^2=D'$, and
$|{\cal T}|^2=D$, 
where
$D'$ and $D$ are the average electron tunneling rates
through the PC barrier 
with and without the presence of the electron in dot 1 respectively. 
Physically, the presence of the electron in dot 1
raises the effective tunneling barrier of the PC
due to electrostatic repulsion.
As a consequence,
the effective tunneling amplitude becomes lower, i.e.,
$D'=|{\cal T}+{\cal X}|^2<D=|{\cal T}|^2$.
This sets a condition on the relative phase $\theta$
between ${\cal X}$ and ${\cal T}$:
$\cos\theta<-|{\cal X}|/(2|{\cal T}|)$.
For simplicity, in this paper we consider the case $\theta=\pi$
as in Refs.\ \cite{Gurvitz97,Korotkov99}.

The unconditional and conditional master equations of the qubit
density matrix, $\rho(t)$, have been obtained and 
written respectively as sets of coupled stochastic
differential equations in terms of the Bloch sphere variables,
$x(t)$, $y(t)$, and $z(t)$, in Ref.\ \cite{Goan00}.
We will use these equations to describe the qubit dynamics.
In the Bloch sphere representation, 
\begin{equation}
\rho(t)=\frac{1}{2}\left (\begin{array}{cc}
1+z(t)&x(t)-iy(t)\\
x(t)+iy(t)& 1-z(t)\end{array}\right ).
\label{bloch}
\end{equation}
Here the variable $z(t)$ represents the
population difference between the two dots. Especially, $z(t)=1$ and
$z(t)=-1$ indicate that the electron is localized in dot 2 and
dot 1 respectively. The value $z(t)=0$ corresponds to an equal
probability for the electron to be in each dot.

\begin{figure}
\centerline{\psfig{file=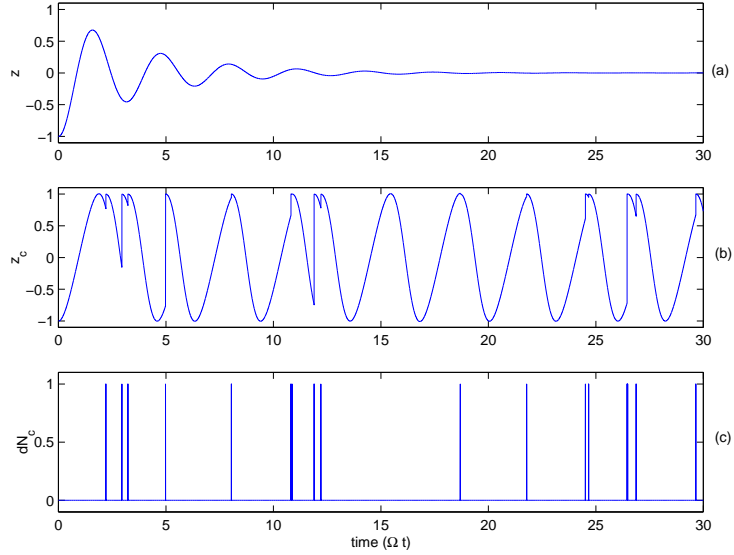,width=0.7\linewidth,angle=0}}
\caption{Illustration for different behaviors between unconditional and
conditional evolutions. The initial qubit state is $|a\rangle$.
The parameters are:  
${\cal E}=0$, $\theta=\pi$, $|{\cal T}|^2=|{\cal X}|^2=2\Gamma_d=\Omega$,
and time is in unit of $\Omega^{-1}$.  
(a) unconditional, ensemble-averaged time evolution of $z(t)$.
(b) conditional evolution of $z_c(t)$. 
(c) randomly distributed moments of detections, which
correspond to the quantum jumps in (b).}
\label{fig:PC-jumps}
\end{figure}

Fig.\ \ref{fig:PC-jumps}(a)
shows the unconditional (ensemble average) time evolution of the
population difference $z(t)$ with the initial qubit state being in
state $|a\rangle$.  
The unconditional population difference $z(t)$, 
rises from $-1$, undergoing some oscillations, and then
tends towards zero, a steady (maximally mixed) state. 
On the other hand, the conditional
time evolution, conditioned on one possible individual realization of
the sequence of measurement results, behaves quite differently. 
We consider first the situation, where 
$D'=|{\cal T}+{\cal X}|^2=0$, discussed in Ref.\ \cite{Gurvitz97}. 
In this case,
due to the electrostatic repulsion generated by the electron, the PC
is blocked (no electron passing through) when dot 1 is
occupied. As a consequence, whenever there is a detection of an electron
tunneling through the PC barrier, the qubit state is collapsed into state
$|b\rangle$, i.e., dot 2 is occupied. 
The conditional, quantum-jump  
evolution of $z_c(t)$ shown in Fig.\ \ref{fig:PC-jumps}(b) 
(using the same parameters and initial
condition as in Fig.\ \ref{fig:PC-jumps}(a)) 
is quite strikingly different from the unconditional one in
that the time evolution is not smooth, but exhibits jumps, and it does
not tend towards a steady state.  One can see that initially the
system starts to undergo a oscillation. As the population
difference grows in time so does the probability for an electron
tunneling through the PC barrier. This oscillation is then interrupted
by the detection of an electron tunneling through the PC barrier,
which bring $z_c(t)$ to the value $1$, i.e., the qubit state is collapsed
into state $|b\rangle$. Then the whole process starts
again.
The randomly distributed moments of detections, $dN_c(t)$,
corresponding to the quantum jumps in Fig.\ \ref{fig:PC-jumps}(b) 
is illustrated in Fig.\ \ref{fig:PC-jumps}(c).
The time evolutions in
Fig.\ \ref{fig:PC-jumps}(a) and (b) 
show little similarity; however, averaging over
many individual realizations shown in (b) leads to a closer and
closer approximation of the ensemble average in (a).

\begin{figure}
\centerline{\psfig{file=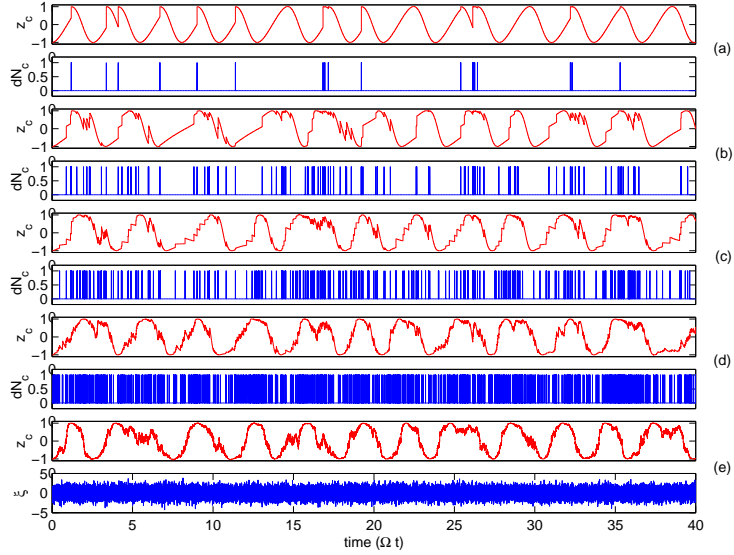,width=0.7\linewidth,angle=0}}
\caption{Transition from quantum jumps to quantum diffusion.
The initial qubit state is $|a\rangle$.
The parameters are:  
${\cal E}=0$, $\theta=\pi$, $|{\cal X}|^2=2\Gamma_d=\Omega$,
and time is in unit of $\Omega^{-1}$.
(a)--(d) are the quantum-jump, conditional evolutions of $z_c(t)$ 
and corresponding detection moments 
with different $|{\cal T}|/|{\cal X}|$ ratios:
(a) 1, (b) 2, (c) 3, (d) 5. With increasing $|{\cal T}|/|{\cal X}|$
ratio, jumps become more frequent but smaller in amplitude.
(e) represents the quantum-diffusive, conditional evolutions of $z_c(t)$.
The variable $\xi(t)$, appearing in the expression of
quantum-diffusive current, 
is a Gaussian white noise with zero mean and unit variance.}
\label{fig:PC-jump-diff}
\end{figure}

Next we illustrate how the
transition from the quantum-jump picture to the quantum-diffusive
picture takes place.
In  Ref.\ \cite{Goan00}, 
we have seen that the quantum-diffusive
equations can be obtained from the quantum-jump description under the
assumption of $|{\cal T}| \gg |{\cal X}|$.  
In Fig.\ \ref{fig:PC-jump-diff}(a)--(d) 
we plot conditional, quantum-jump evolution of $z_c(t)$ and 
the corresponding moments of detections $dN_c(t)$,  
with different $(|{\cal T}|/|{\cal X})|$ ratios.
%equal to  (a) 1, (b) 2, (c) 3, (d)5.  
Each jump (discontinuity) in the $z_c(t)$ curves 
corresponds to the detection
of an electron through the PC barrier.IQC01-preceedings
One can clearly
observes that with increasing $(|{\cal T}|/|{\cal X}|)$ ratio, the
number of jumps increases.
% while their amplitudes decrease.  
The amplitudes of the jumps of $z_c(t)$, however, 
decreases from $D'=0$ with
the certainty of the qubit being in 
state $|b\rangle$ to the case of
$(D-D')<<(D+D')$ with a smaller probability of 
finding the qubit in state $|b\rangle$.
Nevertheless, the population
difference $z_c(t)$ always jumps up
since $D=|{\cal T}|^2>D'=|{\cal T}+{\cal  X}|^2$. 
In other words, whenever there is a detection of an
electron passing through PC, dot 2 is more likely occupied than dot 1.
The case using the quantum-diffusive equations
is plotted in Fig.\ \ref{fig:PC-jump-diff}(e).
We can see that the behavior of $z_c(t)$
for $|{\cal T}|=5|{\cal X}|$ in the
quantum-jump case shown in Fig.\ \ref{fig:PC-jump-diff}(d) is already 
very similar to that of quantum diffusion shown in 
Fig.\ \ref{fig:PC-jump-diff}(e).

%%%%%%%%%%%%%%%%%%%%%%%%%%%%%%%%%%%%%%%%%%%%%%%%%%%%%%%%%%%%%%%%
%%%%%%%%%%%%%%%%%%%%%%%%%%%%%%%%%%%%%%%%%%%%%%%%%%%%%%%%%%%%%
In conclusion,
we have discussed the qubit dynamics and 
illustrated the conditional evolutions by numerical simulations.
The difference between unconditional and
conditional evolutions is demonstrated.
Furthermore, the conditional qubit dynamics evolving
from quantum jumps to quantum diffusion is presented. 
H.-S.G. is grateful for useful discussions with A.~N.~Korotkov,
H.~M.~Wiseman and H.~B.~Sun.

%%%%%%%%%%%%%%%%%%%%%%%%%%%%%%%%%%%%%%%%%%%%%%%%%%%%%%%%%%%%%%%%%%
\vspace{-0.0cm}
%%%%%%%%%%%%%%%%%%%%%%%%%%%%%%%%%%%%%%%%%%%%%%%%%%%%%%%%%%%%%%%%%%%


\begin{thebibliography}{99}
%%%%%%%%%%%%%%%%%%%%%%%%%%%%%%%%%%%%%%%%%%%%%%%%%%%%%%%%%%%%%%%%%%%
\vspace{-0.1cm}
%%%%%%%%%%%%%%%%%%%%%%%%%%%%%%%%%%%%%%%%%%%%%%%%%%%%%%%%%%%%%%%%%%%
%\bibitem[*]{goan} E-mail: goan@physics.uq.edu.au.

\bibitem{Gurvitz97} S.~A.~Gurvitz, Phys. Rev. B {\bf 56}, 15215
        (1997); quant-ph/9808058.

\bibitem{Korotkov99} A.~N.~Korotkov, Phys. Rev. B {\bf 60}, 5737
         (1999); cond-mat/0008461. 

\bibitem{Makhlin98} Y.~Makhlin {\it et al},
        cond-mat/9811029;
        Phys. Rev. Lett. {\bf 85}, 4578 (2000).

\bibitem{Goan00} H.-S. Goan {\it et al},
          cond-mat/0006333, to appear in Phys. Rev. B.

\end{thebibliography}
\end{document}